# A Reconfigurable Chaotic Cavity with Fluorescent Lamps for Microwave Computational Imaging

Ariel Christopher Tondo Yoya, Benjamin Fuchs and Matthieu Davy

Institut d'Electronique et de Télécommunications de Rennes, University of Rennes 1, Rennes 35042, France

**Several computational imaging systems have recently been proposed at microwave and millimeter-wave frequencies enabling a fast and low cost reconstruction of the scattering strength of a scene. The quality of the reconstructed images is directly linked to the degrees of freedom of the system which are the number of uncorrelated radiated patterns that sequentially sample the scene. Frequency diverse antennas such as leaky chaotic cavities and metamaterial apertures take advantage of the spectral decorrelation of transmitted speckle patterns that stems from the reverberation within a medium. We present a reconfigurable chaotic cavity for which the boundary conditions can be tuned by exciting plasma elements, here commercial fluorescent lamps. The interaction of electromagnetic waves with a cold plasma is strongly modified as it is ionized. Instead of being transparent to incident waves, it behaves theoretically as a metallic material. The independent states of the cavity obtained using a differential approach further enhance the degrees of freedom. This relaxes the need of a cavity with a large bandwidth and/or high quality factor. Experimental results validate the use of fluorescent lamps and its limitations are discussed. Images of various metallic objects are provided to illustrate the potentialities of this promising solution.**

## I. Introduction

Computational Imaging (CI) techniques are of great interest to design fast, low-cost and high performances imaging systems.[1-8] Reconstructing images in real-time is especially crucial in the context of security screening,[9,10] through the wall imaging[11] or biomedical imaging[12,13] with electromagnetic waves. The spatial content of a scene is collected onto a set of antenna positions in Synthetic Aperture Radar (SAR) and conventional antenna arrays,[14] whereas in CI it is encoded onto independent spatial patterns generated from only a few sensors. High resolution images are then reconstructed by solving an inverse problem using the knowledge of the sensing matrix between the sensors and the pixels to image.

In acoustics and for electromagnetic waves in microwave range, frequency-diverse antennas including leaky chaotic cavities[15-19] and metamaterial apertures[5, 20-23] exploit the reverberation within a high quality ($Q$) factor medium to generate spatially independent speckle patterns as a function of frequency. A single shot measurement of the broadband impulse responses between the sensors is then sufficient to reconstruct the scene by solving an inverse problem. Thanks to the natural reverberation of waves within the medium, the pattern that illuminates the scene is due to the radiation through the aperture of a spatially random wavefield which is correlated at the scale of a half wavelength. For instance, in chaotic cavities, this random wavefield with Gaussian statistics arises from the superposition of plane waves with random phases. The spatial resolution of the imaging process is governed by the aperture diffraction limit similarly to SAR imaging.

Independent speckle patterns are radiated as the excitation frequency is swept. The spectral correlation length $\delta\omega$ gives the average spectral spacing between two uncorrelated patterns. It is inversely proportional to the reverberation time of the waves within the cavity $\tau$, $\delta\omega = 1/\tau$. The spectral length $\delta\omega$ is related to the $Q$ factor by $Q = \omega_0/\delta\omega$, where $\omega_0$ is the angular central frequency. The number of spectral degrees of freedom $N_\omega$ provided by frequency diverse-antennas is then given by the number of independent states within the working bandwidth $\Delta\omega$, $N_\omega = \Delta\omega/\delta\omega = Q\Delta\omega/\omega_0$. The signal-to-noise ratio (SNR) of the reconstructed images directly depends on $N_\omega$. The same principle is underlying time-reversal focusing within disordered media which makes it possible to focus waves at the diffraction limit with a single antenna.[24]

The reconstruction of an image requires the knowledge of the sensing matrix between the sensors and the scene. To build this matrix, the spectra of radiated field on the scene can for instance be obtained by scanning the near-field of the cavity and propagating this field onto the scene. Achieving high imaging performances requires frequency-diverse antennas with a high number of degrees of freedom. These are antennas with a large bandwidth and/or a high $Q$-factor. However, a trade-off exists between the radiated power toward the scene and the imaging quality. A large aperture indeed increases the radiated power but at the cost of a reduced $Q$-factor due to radiative losses through the aperture.[18] This hence limits the number of spectral degrees of freedom. The SNR is therefore maximized for an optimized aperture.

In this context, another possibility to obtain new degrees of freedom at a single frequency is to change the boundary conditions of the cavity. Those new states can be achieved mechanically and/or electronically. In reverberation chambers, statistically independent measurements are obtained using the rotation of mode-stirrer which consists of metallic reflectors such as paddle wheels.[25] However, the slow rotation of the paddles may be a concern for real-time imaging applications. Electrically tuning the boundary conditions have shown great promise recently. Sleasman *et al.* designed a disordered cavity with a wall replaced by a tunable-impedance boundary for which the reflection phase of each element can be modified using a voltage bias.[17] The unit cell is a resonator with two operating states and the incoming wave is reflected with phase shifts typically close to 0 or $\pi$ over a frequency range of 4 GHz around 20 GHz. A similar tunable metasurface,[26] a so-called spatial microwave modulator (SMM), has been employed to control wave propagation within reverberating media over a bandwidth of 100 MHz.[27,28] The intensity in a reverberation medium can be strongly enhanced at a point by shaping the phase pattern of the SMM so that waves arising from the metasurface interfere constructively at this selected point.[27] In a closed cavity, the SMM also enables to tune the resonances of the cavity in the regime for which the resonance weakly overlap.[28] Nevertheless the bandwidth associated to techniques based on the use of resonators to change boundary conditions is inherently limited since it depends on the linewidth of the resonance.

In this letter, we present a simple and low-cost reconfigurable chaotic cavity for which the boundary conditions are tuned by switching *on* and *off* commercial fluorescent lamps (FL). A FL is a plasma column which is almost transparent to waves when it is not excited but becomes a scattering object as the gas is electrically charged. We use this effect to obtain new states of a chaotic cavity and increase the number of degrees of freedom at a single frequency. Because FLs are lossy objects, we explore the resulting decrease of the $Q$-factor as a function of the number of excited FLs. We then investigate the correlation of speckle patterns for the different states of the cavity. We show that even though the interaction of electromagnetic waves around the plasma frequency with commercial FLs is weak, a differential approach yields an effective number of independent states which greatly enhances the total number of degrees of freedom. We finally demonstrate imaging of metallic objects outside the cavity using computational techniques.

## II. Reconfigurable Chaotic Cavity

The use of plasma tubes to tune the properties of a system in the microwave range is of great interest due to its simple implementation and the large bandwidth that can be considered. The relative dielectric constant of an excited plasma and the angular frequency are theoretically given in the Drude model[29] by $\epsilon_r = 1 - \frac{\omega_p^2}{\omega(\omega+i\nu)}$ and $\omega_p = \sqrt{\frac{n_e e^2}{\epsilon_0 m_e}}$, respectively. Here $n_e$ is the number density of electrons, $\epsilon_0$ is the permittivity of free space, $e$ and $m$ are the electron-charge and electron mass, and $\nu$ is the electron-neutral collision frequency. For $\omega > \omega_p$, an excited plasma has the properties of a dielectric material but at frequencies smaller than the plasma frequency ($\omega < \omega_p$), we obtain $\epsilon_r < 0$ so that the plasma can be considered as a lossy metal. For $\omega_p \gg \nu$, the conductivity $\sigma = \epsilon_0 \omega_p^2/\nu$ of a densely ionized plasma below the plasma frequency is high.

Replacing metallic pieces of traditional antennas by plasma columns makes it possible to design efficient reconfigurable systems.[30] Some successful demonstrations include plasma antennas with low radar cross-sections which are almost undetectable when the plasma is turned off [31], reconfigurable metamaterials,[32-34] tunable leaky-wave antennas,[35, 36] and frequency agile resonators.[37]

Six FLs are located inside an aluminum leaky cavity (see Fig. 1). The cavity of outer dimensions 50x50x30 cm³ is made chaotic by adding three hemispheres inside. Two coaxial to waveguide transitions are attached to two sides of the cavity. The details of the cavity are given in Ref [19]. To change the boundary conditions, the FLs are controlled with an Arduino. $N_s = 64$ states can therefore be achieved. Measurements reported in Ref. [38] of similar commercial FLs in transmission and reflection lead to an estimation of the plasma frequency $f_p = \omega_p/(2\pi)$ around 7 GHz. We choose the frequency range between 7.5 and 8.5 GHz. Even though this is slightly higher than the plasma frequency, this guarantees a cm-resolution since the wavelength at the center frequency $f_c = 8$ GHz is $\lambda \sim$ 3.75 cm.

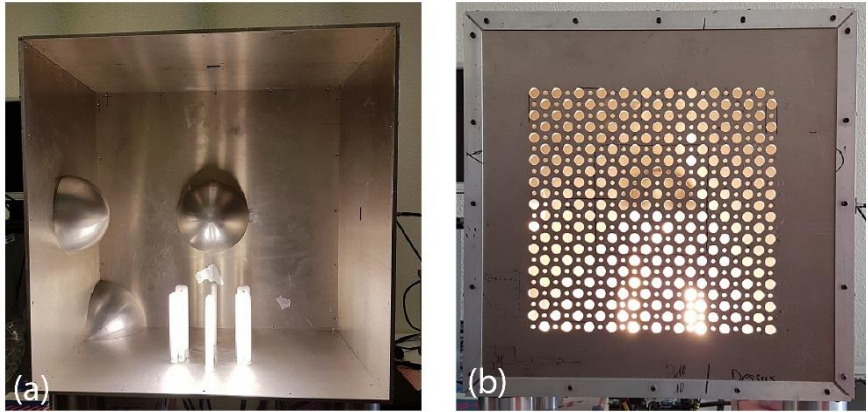

Fig. 1: Photography of the cavity without (a) and with (b) the front face made of drilled holes. The FLs inside can be separately excited using an Arduino.

To illuminate the scene, an array of 11x11 holes of diameter 12.5 mm are perforated on the front face of the cavity. Note that a second array of subwavelength holes with diameter $\lambda/18$ has been drilled but they contribute very weakly to radiations in this frequency range. The near-field of this aperture is scanned on 601 frequency points with an Intermediate Filter bandwidth of 10 kHz on a grid with a spacing of $\lambda/3$ in both sides. For each position of the probe, we measure successively the transmission coefficients for the $N_s$ states. The scanning time of the aperture is about 6h. Spectra of the transmitted intensity through the aperture and their inverse Fourier transform are shown in Fig. 2a for different states. The time signals, which are the responses to 1-ns incident pulses, are spreading over more than 300 ns (see Fig. 2b) due to the reverberation within the cavity. We observe that the intensity decays more rapidly with time when a large number of FLs is excited.

We first explore the $Q$-factor of the cavity as a function of the number of FLs that are excited. To this end, we compute the inverse Fourier transform of the near-field spectra and fit the radiated intensity in the time domain with an exponential decay, $\langle |s(t)|^2 \rangle \sim \exp(-t/\tau)$. $\langle |s(t)|^2 \rangle$ and the exponential decay are shown in Fig. 3a for two states when one and three LFs are turned on, respectively. $Q$ is then found from the formula $Q = 2\pi f_c \tau$. It is seen in Fig. 3b that $Q$ decreases from 3700 to 2000 as a function of the number of excited FLs. The permittivity of the plasma is indeed complex and its imaginary part is responsible of additional losses within the cavity.

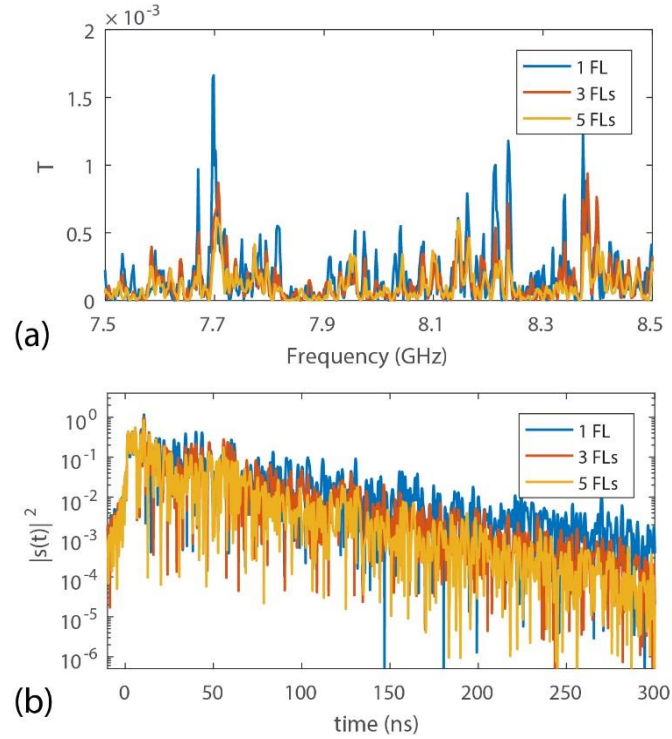

Fig. 2: (a) Spectra of transmission coefficients between the first port of the cavity and a point in the near-field of the cavity for three different states corresponding to one, three and five FLs that are turned on. (b) Transmitted intensity in time-domain for these three different states.

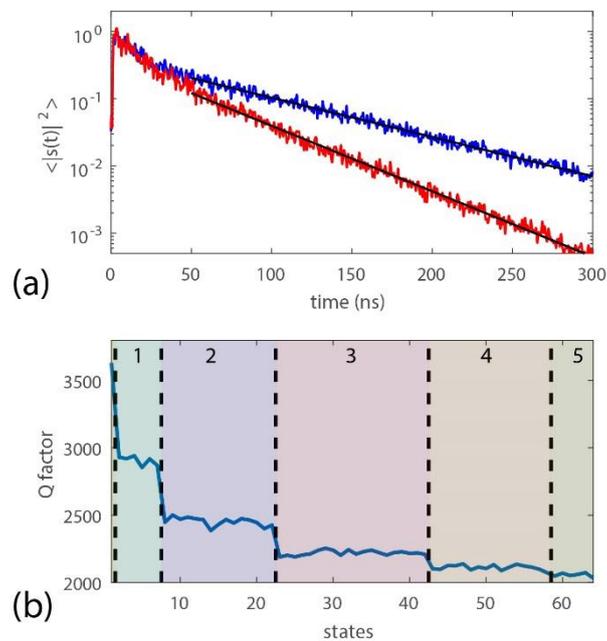

Fig. 3: (a) Plot of the transmitted intensity in time-domain averaged over the near-field of the aperture for 1 (blue line) and 3 (red line) FLs that are turned on. The black lines are exponential fits, $\langle I(t) \rangle = e^{-\frac{t}{\tau}}$. (b) Q factor found from reverberation times $\tau$ as a function of the state. The number of FLs that are turned on is indicated by upper horizontal numbers.

The generation of new configurations of the cavity strongly relaxes the need of a high $Q$-factor for high performances imaging. The total number of degrees of freedom $N$ is theoretically given by the product of the number of transmitting/receiving antennas $N_{\text{ant}}$, the number of spectral degrees of freedom $N_\omega$, and the number of independent states generated at a single frequency by changing the boundary conditions $N_{\text{eff}}$, $N = N_{\text{ant}} N_\omega N_{\text{eff}}$.

To investigate the number of independent states $N_{\text{eff}}$, we compute the correlation coefficient between the spatial vectors of output near-field scans $\phi_i(\omega)$ where the indexes $i$ and $j$ represent the different states. Since the sensing matrix is built from the projection of the near-field scans on the scene, the rank of the covariance matrix built upon those coefficients gives $N_{\text{eff}}$. The correlation coefficient between states $i$ and $j$ is defined as $C_{ij}(\omega) = \left|\phi_j^\dagger(\omega)\phi_i(\omega)\right|/(\|\phi_i(\omega)\|\|\phi_j(\omega)\|)$. The covariance matrix averaged over the frequency range, $\langle C(\omega)\rangle_\omega$, is shown in Fig. 4a. In contrast to the case of perfectly uncorrelated states that would lead to a diagonal covariance matrix, it is seen that the off-diagonal coefficients cannot be neglected. We find $\langle C_{i\neq j}(\omega)\rangle \sim 0.75$. This degree of correlation is also illustrated by the impulse responses for different cavity states. We observe in Fig. 4(c) that those responses are almost identical at short time ($t < 20$ ns) and exhibit significant variations only for $t > 80$ ns.

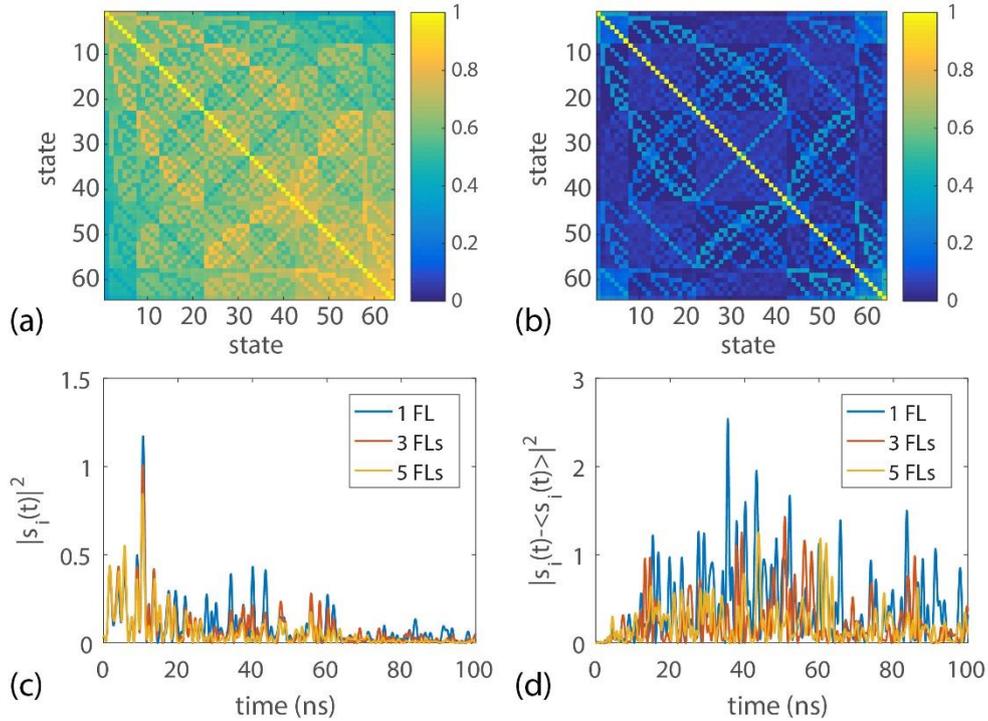

Fig. 4: (a,b) Covariance matrix as a function of the state number averaged over the frequency range for (a) near-field scan and (b) the differential approach (the field averaged over the states is subtracted). (c,d) Plot of the transmitted intensity in the time domain for (c) the near-field scan and (d) the differential approach.

We define the participation number of eigenvalues $\tau_n$ of the covariance matrix as $N_{\text{eff}} = \left(\Sigma_{n=1}^{N_S}\tau_n\right)^2/\Sigma_{n=1}^{N_S}\tau_n^2$. $N_{\text{eff}}$ reflects the distribution of the eigenvalues $\tau_n$ of the covariance matrix. It gives the effective number of independent states that can be generated and therefore the corresponding degrees of freedom of the reconfigurable cavity at a single frequency. In a phase-conjugation experiment, $N_{\text{eff}}$ is equal to the contrast between the focused energy and the background for focusing at a point outside the cavity [39]. Speckle patterns with a vanishing degree of correlation would give a diagonal covariance matrix and $N_{\text{eff}} = N_s = 64$. Here we find $N_{\text{eff}} = 1.5$. The reason of this small $N_{\text{eff}}$ is twofold: 1) the frequency range has been chosen around the plasma frequency and not lower and 2) commercial FLs excited with an almost steady state current are poor reflectors even when $\omega < \omega_p$. It means that the incident field is hence not scattered enough by an excited FL to fully scramble the propagation of waves inside the cavity.

The field radiated from the aperture can be decomposed as the sum of the field that has been transmitted without being scattered by the FL, $\phi_0(\omega)$, and the scattered contribution $\phi_s(\omega)$, $\phi_i(\omega) = \phi_0(\omega) + \phi_s(\omega)$. We estimate the coherent contribution $\phi_0(\omega)$ as the average of the output field over the $N_s$ states, $\phi_0(\omega) \sim \langle \phi_i(\omega) \rangle_i$. For an electrically large chaotic cavity with perfect plasma behaving as transparent/metallic elements with lengths longer than $\lambda/2$, one would expect that $\phi_0$ could be neglected, $\|\phi_0\|^2 \ll \|\phi_s\|^2$. However we find $\langle \|\phi_0\|^2 \rangle = 4.5 \langle \|\phi_s\|^2 \rangle$ which confirms that the scattering coefficient of the plasma tubes is weak. The conductivity of excited FLs is hence likely to be small in comparison to metallic objects. In summary, the propagation of waves within the cavity is only slightly modified by the excitation of the FLs.

A differential approach is proposed to remove the contribution that is robust to the average over the states of the cavity and we consider the so-corrected patterns $\tilde{\phi}_i(\omega) = \phi_i(\omega) - \langle \phi_i(\omega) \rangle_i$. The covariance matrix computed from $\tilde{\phi}_i(\omega)$ is seen in Fig. 4b to be nearly diagonal. $N_{\text{eff}}$ is then greatly enhanced and found to be 11.5. This procedure mainly removes the coherent field $\phi_0$ and strong variations between time signals for different states are now seen in Fig. 4(d) even at short times.

To assess the impact of the FLs on the number of degrees of freedom of the cavity, we compute the singular value decomposition of the sensing matrix of the cavity. In Fig. 5, the distribution of normalized singular values $\tilde{\lambda}_n = \lambda_n/\lambda_1$ is shown for $N_\omega=61$ and different numbers of states that are randomly chosen over the 64 available states. For $n < 205$, it is seen that the distribution of $\tilde{\lambda}_n$ is more flat as $N_s$ increases. This indicates that the number of uncorrelated speckle patterns also increases. However, we observe that the slope saturates as $N_s > 15$. It is possible to select appropriately only a few states among the 64 in order to maximize $N_{\text{eff}}$. By using a standard genetic algorithm, we have selected only 12 states and managed to achieve $N_{\text{eff}} = 7.8$. The total degrees of freedom is then $N = N_\omega N_{\text{eff}} = 476$. In comparison, the cavity without the FLs exhibits a Q-factor that is approximately twice higher leading to an estimate number of degrees of freedom of $N = N_\omega = 128$.

Furthermore, in all cases, this distribution falls rapidly for $n > 205$. The number of significant singular values is indeed bounded due to the dimensions of the aperture $A = 0.09$ m$^2$. The maximum number of independent states is $A/(\lambda/2)^2$. Here we find $A/(\lambda/2)^2 = 225$, where $\lambda$ is the wavelength corresponding to the smallest frequency, which is in good agreement with the index of the break in the slope.

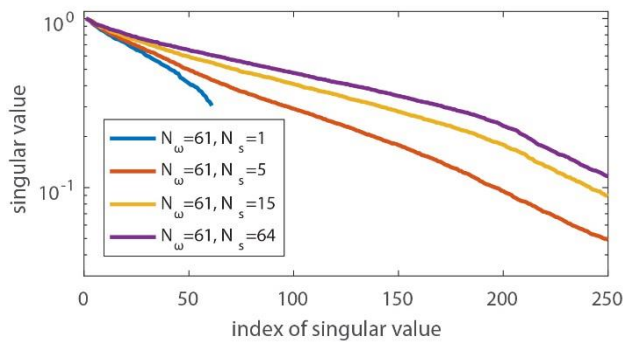

Fig. 5: Variation of the normalized singular values of the sensing matrix of the cavity with their index for various FL configurations.

### III. Imaging Results

The scene is illuminated by a horn antenna located 40 cm away from the center of the cavity and the back-scattered signal is recorded by the port of the cavity. We assume the first order Born approximation so that the received signal, for a cavity at state *i*, can be written as[18]

$$y_i(\omega) = \int dr G_0(r_1, r, \omega) \Sigma_n G_0(r, r_n, \omega) x(r) h_i(r_n, \omega). \quad (1)$$

Here $G_0$ is the free-space Green's function, $r_1$ is the location of the horn antenna, $x(r)$ is the reflectivity of the scene at point $r$ and $h_i(r_n, \omega)$ is the response between location $r_n$ in the near-field of the cavity in its $i^{\text{th}}$ state and the port. To reconstruct $x(r)$, we take into account that the subtraction of the signal averaged over the states of the cavity greatly enhances the degrees of freedom at a single frequency and instead solve the following discretized equation

$$\mathbf{y} - \langle \mathbf{y} \rangle_i = (\mathbf{H} - \langle \mathbf{H} \rangle_i) \mathbf{x}. \quad (2)$$

Here $\mathbf{y}$ is the measurement vector including all the frequencies and all the cavity states, $\mathbf{H}$ is the corresponding sensing matrix between the emission/reception and the pixels at which $\mathbf{x}$ is reconstructed. We subtract at each frequency the average values over the cavity states: $\langle \mathbf{y} \rangle_i$ for the observation vector and $\langle \mathbf{H} \rangle_i$ for the sensing matrix.

We discretize the scene with square pixels of dimensions $\lambda/6 \times \lambda/6$ and solve Eq. (1) to reconstruct objects located at a distance $z = 52$ cm from the aperture of the cavity. We use $N_\omega = 61$ frequencies and $N_s = 12$ states. The images in the plane of objects of a sphere, two spheres, a metallic bottle and a L-shaped phantom are shown in Fig. 6. To invert the matrix $(\mathbf{H} - \langle \mathbf{H} \rangle_i)$, we first perform a truncated singular value decomposition and then apply a total variation norm regularizer.[40] The SNR of the images defined as the maximum of the target over the maximum of the background are 33 dB, 28 dB, 16 dB and 8 dB respectively. A 3D reconstruction of the two spheres is also shown in Fig. 6e. The resolution in range is now given by $c_0/B$, where $B$ is the bandwidth. Here the $N_\omega$ frequencies are chosen to span the frequency range so that $B = 1$ GHz, giving a theoretical resolution of 3cm along the z-direction.

The good reconstruction of the shape of the objects demonstrates the capability of our system to image metallic objects with a resolution of $\sim 5$ cm in the (x-y) plane. This also confirms the validity of the differential approach for computational imaging with our reconfigurable cavity.

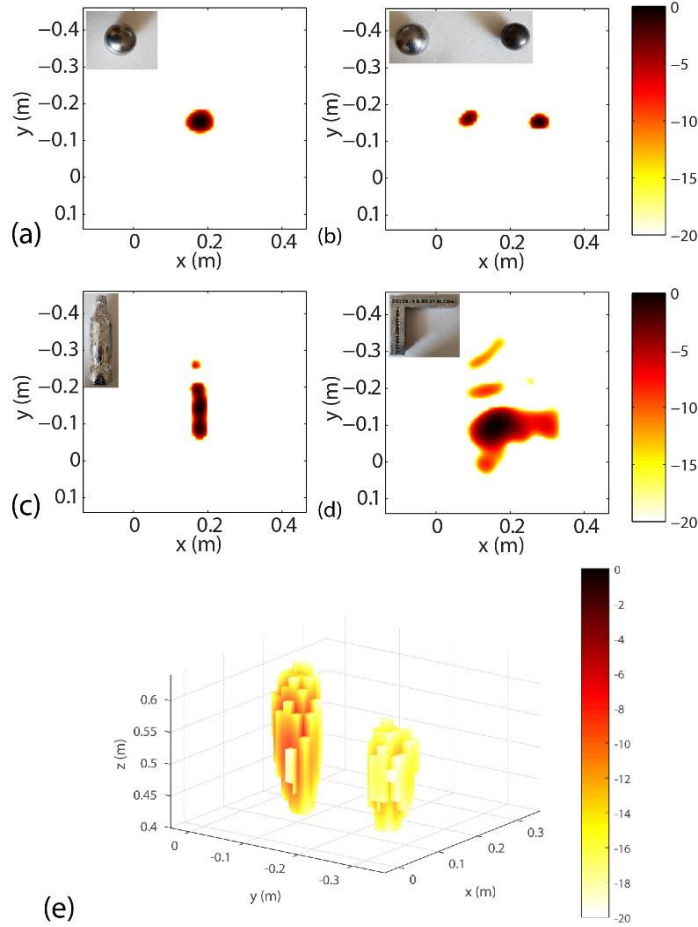

Fig. 6: (a) Reconstructed images of 4 objects at a distance of 52 cm from the aperture of the leaky cavity with $N_f$=61 and $N_s$=12: a single (a) and two spheres (b) of 5-cm diameter; (c) a metallic bottle of height 25 cm and diameter 9 cm; and (d) a metallic L-shaped phantom. (e) 3D reconstruction of the two spheres shown in (b).

## IV. Conclusion

We have demonstrated the design of a reconfigurable leaky cavity for microwave computation imaging using commercial FLs. Thanks to the remarkable properties of cold plasma regarding the interaction with microwaves below the plasma frequency, new states of the leaky cavity can be generated at a single frequency by exciting plasma columns. This strongly increases the available degrees of freedom that are crucial to reconstruct a scene using computational techniques. At a given SNR, introducing new states by changing the boundary conditions of frequency diverse antennas enables to reduce the frequency range of measurements and/or consider new systems with smaller $Q$ factor. In contrast to tunable metasurfaces designed using resonators, the frequency range over which new states of the cavity can be generated in principle spans the complete frequency range below the plasma frequency.

Nevertheless we have shown that an excited commercial FL is far from behaving as a metallic object. The reflection of waves at its interface is weak and this leads to a small number of independent states and a reduced $Q$ factor. However, we have considered a differential approach to remove the coherent part of the field and achieve a high effective number of independent states. Our low-cost system, that is intended to be a proof of concept, could be largely improved by using more sophisticated plasma columns with higher electron density to 1) reduce the losses of the excited columns; and 2) increase the plasma frequency so that higher frequency ranges can be considered. Measurements of efficiencies up to 50% have been reported for plasma column elements in comparison to similar metallic objects [31]. Additional independent states could therefore be obtained without resorting to the differential approach and this would further improve the SNR of the imaging process.

**Acknowledgment**

The authors would like to thank Frédéric Boutet and Jean-Christophe Le Cun for their help in setting up the experiment.